# Plasmonic Gas Sensing based on Cavity-Coupled Metallic Nanoparticles

Jian Qin[1], Yu-Hui Chen[2], Boyang Ding[2]*
Richard J. Blaikie[2*], and Min Qiu[1]*

[1]*State Key Laboratory of Modern Optical Instrumentation, Department of Optical Engineering, Zhejiang University, Hangzhou 310027, China*
[2] *MacDiarmid Institute for Advanced Materials and Nanotechnology, Department of Physics, University of Otago, PO Box 56, Dunedin 9016, New Zealand*

\* boyang.ding@otago.ac.nz
\* richard.blaikie@otago.ac.nz
\* minqiu@zju.edu.cn

**Abstract:**
Here we demonstrate the gas sensing ability of cavity-coupled metallic nanoparticle systems, comprising gold nanoparticles separated from a gold mirror with a polymer spacer. An increase in relative humidity (RH) causes the spacer to expand, which induces a significant reduction of nanoparticle scattering intensity, as the scattering is highly dependent on the cavity-nanoparticle coupling that closely relates to the nanoparticle-mirror distance. This lithography-free structure enables a remarkable averaging sensitivity at 0.12 dB/% RH and 0.25 dB/% RH over RH range (45-75%), possessing an estimated resolution better than 0.5% RH with full reversibility and almost zero-hysteresis, exhibiting notable gas sensing potentials.

In metallic nanoparticles (NPs), collective oscillation of conduction electrons can be excited upon light illumination. This is known as the excitation of localized surface plasmons (LSPs), giving rise to the concentration of incident light into deep subwavelength volumes. As the result, the electromagnetic (EM) fields can be significantly enhanced, leading to the intensified interactions to neighboring matters. Based on this property, in the past decade, extensive effort has been paid to develop gas or vapour sensing applications using metallic NPs, as this allows for sensitive, remote and non-invasive optical read-out in the nearby presence of gas or vapour molecules [1].

For example, the extinction spectra of gold (Au) or silver (Ag) NPs can be modified when the NPs are soaked in helium and argon gases [2], as the refractive index change of surrounding air can induce a resonance shift of LSPs. If the NPs are made of materials that are catalytically active to certain gases, e.g. palladium can absorb hydrogen gas forming palladium hydride, the resonance shift of LSPs can be significantly enhanced [3–7]. However the sensitivity is highly limited by the large LSP linewidth that is induced by strong radiative damping of metals. In order to improve the sensitivity, dimer NP structures [8] or periodic arrays of metallic NPs [9–11] have been lithographically constructed to create EM hot spots or enhance the diffractive coupling of LSPs, aiming to sharpen the plasmonic resonances.

Another solution is to couple metallic NPs with optical cavities, e.g. placing NPs above a mirror. A proper distance from the mirror allows for constructive interference between the scattered light from NPs and the reflected light from the mirror. Without the need for complicated and expensive lithographic techniques, the cavity-NP coupling can greatly reduce the linewidth [12], and also highly enhance the LSP resonances [13] [14], enabling a wide range of applications, e.g. photoemission enhancement [15], surface enhanced Raman scattering [16–18], bio- or chemical sensing [12,19–22] and temperature sensing [23]. Recently Wirth et al discovered that tuning the position of individual NPs with respect to the mirror can effectively alter the cavity-NP coupling states, leading to the spectral and directional reshaping of nanoparticle scattering [24]; while we found that this position tuning can also induce a large modification of single NP scattering intensity [25].

Here we report a novel gas sensing regime utilising this cavity-coupled single NP system based on an example of humidity detection. As shown in Fig. 1, AuNPs are separated from a Au mirror with a Polyvinyl alcohol (PVA) film, which can swell/de-swell upon the absorption/desorption of ambient water molecules [26], causing the variation of film thickness or namely the NP-film distance. As revealed by measured far-field patterns and simulated near-field distributions, the change in NP-film distance can effectively alter the cavity-NP coupling state, thus leading to the modification of scattering intensity from individual NPs. In particular, when relative humidity (RH) changes from 5% to 75%, the scattering intensity of a single NP reduces down to ~1/8 of its original value, resulting in a remarkable averaging sensitivity of ~0.12 dB/% RH and an estimated resolution better than 0.5% RH. In addition, the great RH sensitivity exhibits zero-hysteresis and full reversibility with a rapid response time, manifesting the outstanding gas sensing potentials of the cavity-coupled NP system.

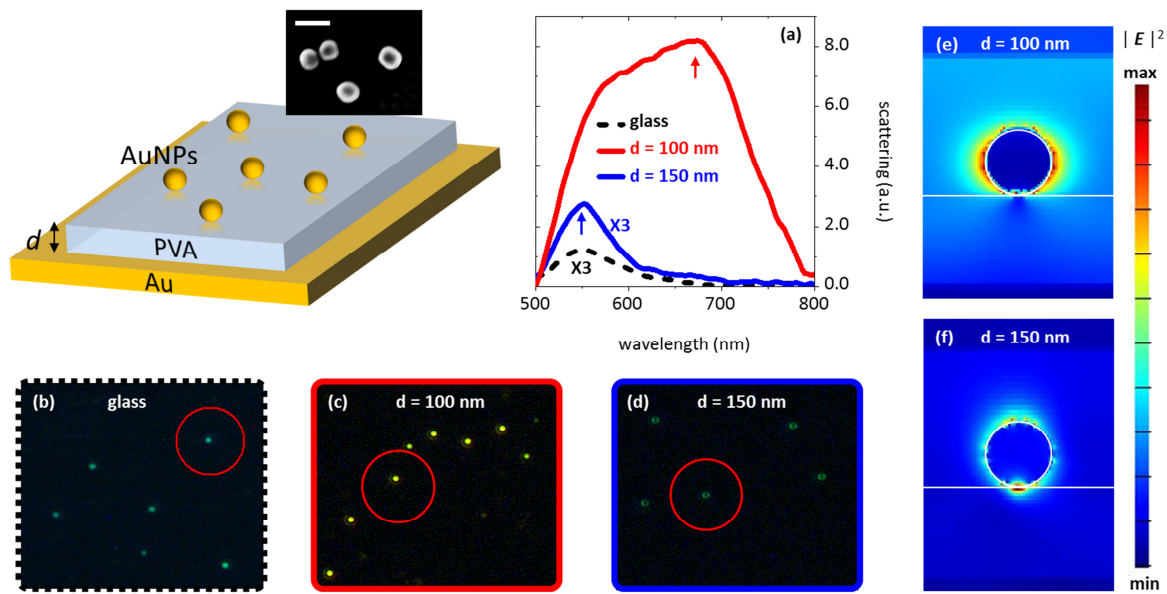

**Figure 1** Structure description and optical properties. (a) Measured scattering spectra of individual AuNPs on glass substrate (dashed), and $d$ = 100 nm (red) and $d$ = 150 nm (blue) above a Au substrate (scattering for AuNPs on glass and 150 nm from the Au film were multiplied by 3); while panel (b), (c) and (d) show far-field images of these AuNPs (marked by red circles) on different substrates, respectively. Note: The white scale bar in the SEM image of the schematic stands for 100 nm; while all the optical measurements were performed at RH = 35%. Panel (e) and (f) show the electric field intensity |$E$|[2] of AuNPs placed 100 and 150 nm above the Au mirror, respectively.

In particular, a ~100 nm thick Au film has been thermally evaporated onto a silicon substrate,

followed by the deposition of a PVA film using a spin-coater. Adjusting the concentration of PVA solution and spinning speed, the PVA film thickness can be tuned and was finally confirmed by a surface profiler (Model:) at RH = 35% and room temperature, with a precision of ±5 nm. Spherical AuNPs with diameter of $\phi = 60 \pm 7$ nm were chemically prepared using the method [27,28] and stored in ethanol suspension, while their physical appearances are characterised by a scanning electron microscope (SEM) as shown in the inset of the schematic of Fig. 1. Finally, these AuNPs were placed on the top of PVA spacers. A dark-field microscope associated with a gas chamber was used to collect the scattering spectrum and far-field images from single NPs. The RH in the gas chamber was adjusted by being ventilated with mixtures of dry and humid nitrogen and was recorded every 1 second with an accuracy of ±5 % RH using a commercial electrochemical RH sensor (Thorlabs TSP01). Finite-Difference-Time-Domain (FDTD) methods (Lumerical Solutions) were used to simulate the near-field distributions and optical spectra of the structures, while the refractive index of Au and PVA within in different humidity environments were acquired from Ref [29] and [30], respectively.

Turning into details, Fig. 1(a) shows that the scattering spectrum of a AuNP on a glass substrate (black dashed curve) peaks at 554 nm. In contrast, when AuNPs are placed $d = 100$ nm above a Au film (red curve in Fig. 1a), they gain considerably enhanced scattering intensity (up to 20 times compared to those on a glass substrate) and a much wider linewidth with a maximum at $\lambda = 674$ nm, indicating an overlap of multiple resonances [31]. If the NP-film distance increases to $d = 150$ nm (blue curve), the scattering maximum blue-shifts back to $\lambda \approx 550$ nm with a large intensity reduction. Far-field scattering patterns also exhibit a high degree of dependence on substrates. For example, the scattering image of AuNPs on glass (Fig. 1b) display dot-shaped green patterns, while NPs placed $d = 100$ nm above the Au film (Fig. 1c) show much brighter yellow dots. It is noted that the scattering of AuNPs on a 150 nm spacer (Fig. 1d) exhibits green doughnut-shaped patterns, rather than the dots pattern as shown in AuNPs supported by other substrates. As revealed by Ref [24] and our previous study [25], these substrate-induced AuNPs' optical properties are the result of resonance coupling between LSPs and optical cavity modes.

The FDTD simulations of near-field distributions demonstrate such coupling effects. In particular, the simulated field intensity $|E|^2$ of the AuNP placed 100 nm above the Au film was shown in Fig. 1(e). This field distribution corresponds to the scattering maximum at $\lambda = 674$ nm that is indicated by the red arrow in Fig. 1(a), demonstrating a clear resonance horizontally oriented with respect to the Au film. In contrast, $|E|^2$ at $\lambda = 560$ nm (indicated by the blue arrow in Fig. 1a) of the AuNP above a 150 nm spacer shows a vertically oriented resonance. These results highly correspond to the far-field observations of dot and doughnut shape patterns. It is apparent that varying the NP-film distance can effectively tune the coupling state between AuNPs and the optical cavity, giving rise to the change of near-field dipole orientation. As the results, far-field patterns are also modified as the NP-film distance changes.

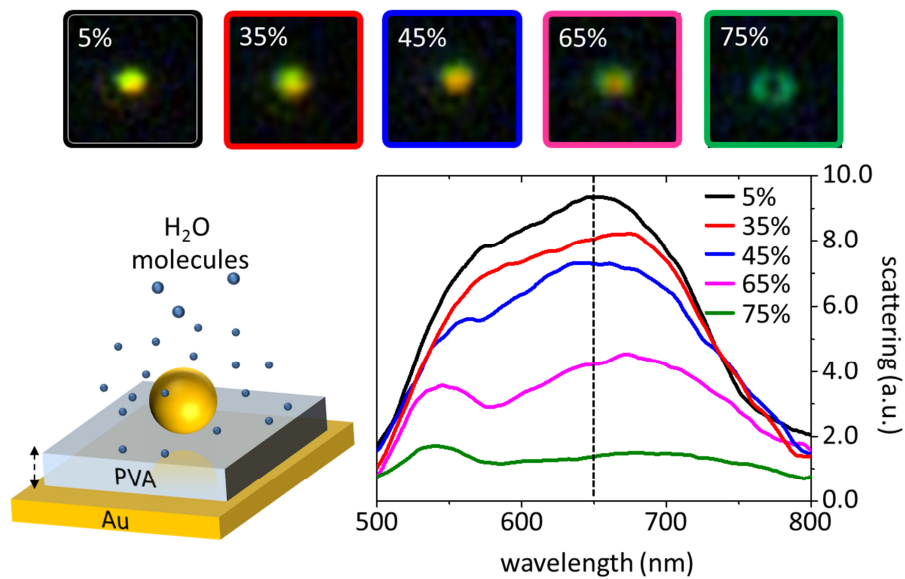

**Figure 2** Optical properties variation upon interaction with water molecules. Scattering spectra of a AuNP separated from the Au mirror with a PVA spacer measured with RH = 5% (black), 35% (red), 45% (blue), 65% (magenta) and 75% (green). The corresponding far-field images are shown above. The thickness of the PVA film is 100 nm measured at RH = 35%. The dashed line indicates the wavelength of $\lambda = 650$ nm.

Next we will demonstrate the humidity sensing capability of the cavity-coupled NP structures.

Specifically the scattering spectrum and far-field image of a AuNP separated from a Au film with a PVA spacer will be measured within different humidity environments. In this experiment, the initial thickness of the PVA spacer was confirmed as $d = 100$ nm at ambient RH = 35%. As shown in Fig. 2, the scattering spectrum at RH = 5% (black curve) shows a broad linewidth with maximum at $\lambda = 650$ nm. As the humidity increases, the scattering magnitude at $\lambda = 650$ nm drastically reduces, while another maximum at $\lambda \approx 550$ nm gradually appears. When RH achieves at 75% (green curve), the long wavelength band maximum almost disappears while there is only one maximum at $\lambda = 545$ nm, highly resembling the scattering spectrum of the AuNP placed on a $d = 150$ nm spacer (blue curve in Fig. 1a). The far-field observation shows a similar tendency that the scattering image captured at RH = 75% (green framed image in Fig. 2) displays a green doughnut pattern, which is identical to the far-field image of AuNPs 150 nm above the mirror, as shown in Fig. 1(d). These evidences unambiguously indicate that the humidity induced modification of spectra and far-field patterns are the result of NP-film distance variation, as absorption/desorption of water molecules leads to the expansion/shrinking of the PVA film.

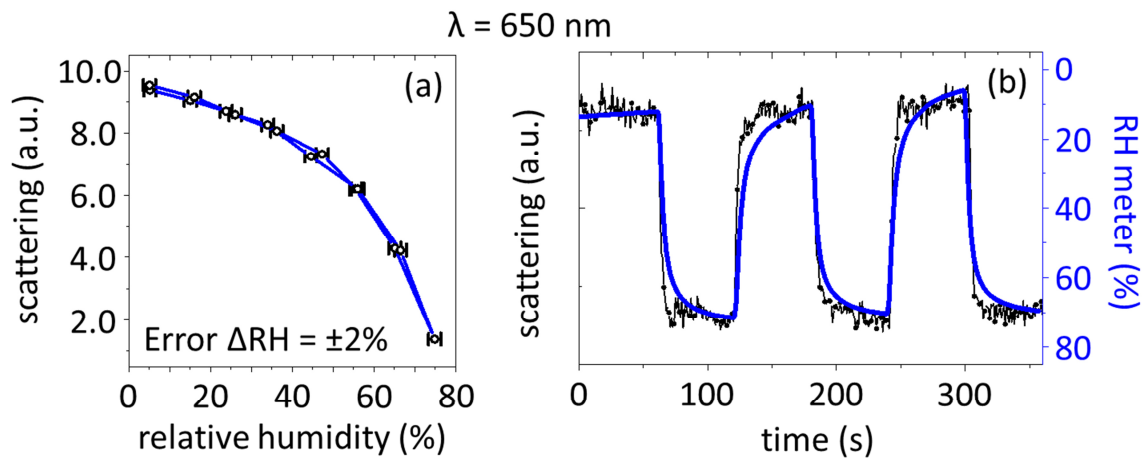

**Figure 3** Sensing performance of the cavity-coupled NP structure. (a) the scattering intensity of the cavity-coupled AuNP (the same one as in Fig. 2) measured at $\lambda = 650$ nm with increasing and decreasing humidity. (b) the scattering intensity (black dot-line curve) varies as a function time when humidity cycles between 5% and 70%, as compared to readings (blue curve) from a commercial electrochemical sensor.

Fig. 3(a) shows the scattering intensity at $\lambda = 650$ nm of the single AuNP above a 100 nm PVA spacer with different RH values changing from 5% to 75% and then back to 5%. First of all, we note that the scattering magnitudes are almost identical for RH variation along opposite directions, indicating that the optical response of the cavity-coupled NP system to RH changes is fully reversible and has almost no sensing hysteresis within the measuring error ±2% RH. In addition, as the result of modified cavity-NP coupling, the scattering intensity at RH = 75% is only ~1/8 of that at RH = 5%. We note that such a large humidity-induced intensity reduction allows for an averaging RH sensitivity of 0.12 dB/% RH over a wide RH range (5%-75%), which is a remarkable sensitivity improvement as compared to other nanophotonic humidity sensors, e.g. 0.05 dB/% RH of the dye-doped ultrathin-film sensor [26,32], 0.07 dB/% RH of the nanofibre-based plasmonic sensor [33] and 0.09 dB/% RH of the plasmonic gap resonance sensor [34]. Moreover, it is worth noting that since the scattering intensity of the cavity-coupled NP system shows a nonlinear change over the whole RH variation, at certain RH ranges the sensitivity can be even higher. For example, for the RH range from 45% to 75%, the sensitivity of the system achieves as high as ~0.25 dB/% RH. Given this exceptionally good sensitivity and the photon detection resolution of our spectrometer, we estimate the sensing resolution of the cavity-coupled NP system to be better than 0.5% RH, which is 4~5 times higher than that of most commercially available electrochemical sensors.

Next we directly compare our cavity-coupled NP system with the electrochemical sensor used in our experiment. Fig. 3(b) shows the scattering intensity of a single AuNP above a 100 nm spacer (black dot-dashed line) changes as a function of time when RH values cycles between 5% and 70%; while the simultaneous readings from the electrochemical sensor (blue curve) are plotted as a reference. We note that the scattering variation mostly overlaps with the sensor readings. However when the humidity suddenly changes, e.g. at $t = 70$ s, RH abruptly jumps from 5% to 70%, the scattering intensity of the AuNP shows an immediate response, while the sensor readings change slowly, taking ~30 s to achieve accurate values. Similar behaviours can be observed at all time points when RH suddenly jerks. Therefore we can conclude that the cavity-coupled NP system gains way more rapid response than does the commercial electrochemical sensor, with less than 1 s response time when RH changes from 5% to 70%. We expect to see a much shorter response time if the experiment can be performed using sensors with finer time resolution, as PVA films exhibit much faster humidity response in other studies [35]. Moreover, Fig. 3(b) also demonstrates excellent reversibility and repeatability of the cavity-coupled NP system; while it is noteworthy that our system can achieve full reversibility without extra thermal treatment as used in some previous studies [36].

In conclusion, we have demonstrated humidity sensing based on cavity-coupled metallic nanoparticle systems. Specifically, being separated from a gold film with a PVA spacer, the scattering of AuNPs can be dramatically modified when ambient humidity changes, as the PVA spacer can expand or shrink upon interaction with water molecules, causing the NP-film distance variation. FDTD simulations and far-field patterns manifest that the NP-film distance change can effectively tune the cavity-NP coupling state, thus enabling the large modification of scattering intensity. This LSP resonance based structure offers high averaging sensitivity of 0.12 dB/% RH and ~0.25dB/% RH over the RH range 45% to 75% with an estimated resolution better than 0.5% RH. In addition, this system exhibits full reversibility and repeatability with rapid response time and almost no hysteresis. Together with the lithography-free fabrication process, the remarkable sensing performance of the cavity-coupled NP system provides a versatile platform towards various gas sensing applications, as long as proper sensing materials can be adopted to replace PVA films [36].